\documentclass[11pt]{article}
\pdfoutput=1 
\usepackage[pdftex]{graphicx}

\begin{document}

\title{
\large
A Simple Redistribution Vortex Method (with Accurate Body Forces)}
\small
\author{O.R.\ Tutty}
\date{School of Engineering Sciences\\
University of Southampton\\
Southampton SO17 1BJ, UK\\
}
\maketitle


\begin{abstract}
A circulation redistribution scheme for viscous flow is presented.
Unlike other redistribution methods, it operates by transferring the circulation 
to a set of fixed nodes rather than neighbouring vortex elements.  A new 
distribution of vortex elements can then be constructed from the circulation 
on the nodes.  
The solution to the redistribution problem can be written explicitly as a set 
of algebraic formulae, producing a method which is simple and efficient.  
The scheme works with the circulation contained in the vortex elements only, 
and does not require overlap of vortex cores.  With a body fitted redistribution 
mesh, smooth and accurate estimates of the pointwise surface forces 
can be obtained.  The ability of the scheme to produce high resolution 
solution is demonstrated using a series of test problems.  
\end{abstract}


\vspace{0.5cm}
\section{Introduction}

Discrete Vortex Methods (DVM's) are Lagrangian methods for solving for 
rotational fluid flow in which the vorticity field is partitioned into 
a finite number of discrete vortex elements, and the evolution of 
the flow field is determined by following the motion of the vortex elements. 
One advantage of DVM's is that the computational effort is applied only in the 
regions of most interest.  Also, for flow past bodies, the far field 
conditions are automatically satisfied as the vorticity field will decay 
to zero away from the body, avoiding the problems that can occur when 
truncating the domain in grid based methods.

For two-dimensional inviscid flow, there is a single component of vorticity, and 
for inviscid flow, the elements are convected with constant strength at the 
local velocity of the fluid.  For viscous flow, some method of modelling 
the viscous diffusion must to added to the numerical scheme.  A number of 
schemes have been developed.  One of the first applications of a DVM for 
viscous flow \cite{Chorin} used a random walk to model the 
viscous effects.  A random walk is simple to apply but has relatively low 
resolution and produces noisy results.   A more sophisticated method 
is the Particle Strength Exchange (PSE) method, introduced in \cite{DMG}, 
which models the viscous effects using integral operators.   A related method is the vorticity 
redistribution method of Shankar and van Dommelen \cite{SvD,Shankar} which, like PSE, 
involves redistributing circulation between the elements, but does not 
require a regular mesh, introducing elements locally as required.  
A more recent redistribution. based on time dependent Gaussian cores, is 
given by \cite{LG}.  
A description of  many aspects of vortex methods  
can be found in the book by Cottet and Koumoutsakos \cite{Cottet}.  Also, 
a comparison of four viscous methods which do not require the introduction 
of a grid can be found in \cite{KT}.

Many papers on vortex use the total force (coefficients) to assess the 
accuracy of the method but do not present details of the pointwise surface 
forces.  These are frequently noisy due to the irregular distribution of 
vortex particles, as can be seen in \cite{SS}, who use random walk 
for the diffusion.  In fact, the total force can also be noisy and require 
smoothing, as in \cite{SS,CT}.  In one of the benchmark papers using vortex 
methods, Koumoutsakos and Leonard \cite{KL} do present plots of the 
surface pressure and vorticity, although they observe 
high frequency oscillations in the surface vorticity.

Koumoutsakos and Leonard \cite{KL} use the PSE method for diffusion. PSE involves 
the introduction of a grid to the scheme with frequent remeshing in order to 
maintain the accuracy of the calculation.  It appears that the order 
introduced into the method by the remeshing helps smooth out the 
spurious high frequency oscillation in the forces that are found in some other 
methods.  Hence, a possible way of reducing the noise would be to remesh 
every time step with a PSE method.  However, an alternative is to combine the 
remeshing and diffusion operation in single step.  This can be done by 
making the redistribution of the circulation satisfy certain constraints 
which have a physical meaning.  This is a variation of the redistribution 
method of Shankar and van Dommelen \cite{SvD}, but rather than redistributing 
circulation between vortex elements, the circulation contained in 
each discrete vortex element is independently transferred onto a small 
set of neighbouring nodes .  The solution for this redistribution 
problem can be written 
explicitly as a set of algebraic equations.  The result is a simple, 
efficient scheme which maintains a regular distribution of vortex elements.  
The surface force can be calculated directly from local variables, showing 
very good agreement with test data.  Also, since this scheme works with 
the circulation of a vortex element, independent of the vorticity 
distribution in the element, there is no requirement for overlap of cores, 
as in the PSE method.

The paper is structured as follows; first, an outline of the basic method 
is presented.  This is followed by a description of the redistribution scheme.    
Means of calculating the body forces are then discussed.  A brief description of 
the finite-spectral code used to generate test data is given.  
Test results are then presented for impulsively started flow past a cylinder and 
a square.  Finally, some conclusions are given.

\vspace{0.5cm}
\section{Basic Method}

The two-dimensional incompressible Navier-Stokes equation in vorticity form is 
\begin{equation}
{D \omega \over Dt}~=~ {\partial \omega \over \partial t} \,+\, 
u\,{\partial \omega \over \partial x} \,+\, v\,{\partial \omega \over \partial y} 
~=~ \frac{1}{Re}\nabla^2 \omega
\label{NS}
\end{equation}
where $(u,v)$ are the velocity components in Cartesian coordinates $(x,y)$, 
$t$ is the time, 
$\omega$ the 
vorticity, $Re$ the Reynolds number, and $\nabla^2$ the two-dimensional 
Laplace operator.  The velocity is normalised by the  free stream velocity 
$U_0$, the length scales by a characteristic length $L$, and time by 
$L/U_0$.  The Reynolds number is given by $\rho L U_0/\mu$ with 
$\rho$ and fluid density and $\mu$ its dynamic viscosity.  

Consider an individual vortex element centred on $z=z_j$ where 
$z=x+i\,y$ gives the coordinates in complex form, with a vorticity 
distribution $\gamma(\eta)$ where $\eta=| z-z_j |$, where  
$2\pi\int_0^\infty \gamma(\eta)\eta\,d\eta=1$ so that the distribution 
function has unit circulation.   
The vorticity field is represented by $N$ discrete vortex elements so that 
\begin{equation}
\omega~=~\sum_{j=1}^N\,\Gamma_j\,\gamma(|z-z_j|)
\label{V2}
\end{equation}
where $\Gamma_j$ is the strength of vortex $j$.

The velocity generated by the vortex elements is given by
\begin{equation}
u_b+i\,v_b~=~\sum_{j=1}^N\,i\, \Gamma_j{z-z_j \over |z-z_j|^2}F(|z-z_j|)
\label{VV}
\end{equation}
where 
\begin{equation}
F(\eta)~=~\int_0^\eta \gamma(s)\, s\,ds
\label{VV_k}
\end{equation}

A number of different functions can be used for the vorticity distribution 
$\gamma$.  In general, point vortices given by a delta function 
are not used.  Instead, a smooth distribution which does not have the numerical 
problems associated with point vortices is adopted.  
A standard distribution, used here, is the Gaussian vortex,  
\begin{equation}
\gamma(\eta)~=~\frac{1}{\pi\sigma^2}\,e^{-\eta^2/\sigma^2}
\label{LV}
\end{equation}
where $\sigma$ is a measure of the core size of a vortex.  
This gives 
\begin{equation}
F(\eta)~=~\frac{1}{2\pi}\left[ 1-e^{-\eta^2/\sigma^2}\right]
\label{LVV}
\end{equation}

As usual, the boundary conditions at the surface of the body are satisfied 
by the use of a vortex panel method.  
Both standard straight, constant strength, vortex panels and 
the higher order panels given in \cite{CT} \footnote{The vortex panel method 
presented in \cite{CT} 
contains a typographical error.  The formula for the velocity 
generated by a panel ($\mathbf{u}^*(z)$ in equation 16) is from a definite integral 
and should be the difference between the two terms not the sum of them.} 
were investigated.   
The latter method uses overlapping curved panels with a linear distribution 
of vorticity along the panels.  This allows an accurate 
representation of a smooth body and produces a  velocity distribution 
which is singularity free.  However,   
there was little difference in the results for the different panels for 
flow past a circular cylinder if enough panels were used.  
The cylinder results (Section \ref{cylinder_results}) presented below 
use the high order panels, 
but, because of the sharp corners,  
constant strength, straight panels were used for flow past a square 
(Section \ref{square_results}).  

The velocity now consists of three components, 
\begin{equation}
{\mathbf u}={\mathbf U}_f+{\mathbf u}_b+{\mathbf u}_p
\label{vel_tot}
\end{equation}
where ${\mathbf u}=u+i\,v$ is the fluid velocity, 
${\mathbf U}_f=U_0$ is the free stream 
velocity, ${\mathbf u}_b=u_b+i\,v_b$ is the velocity generated by 
the vortex elements, and ${\mathbf u}_p=u_p+i\,v_p$ generated by the panels used 
to satisfy the boundary conditions at the surface of the body.

Numerically, an operator splitting method is used, with inviscid 
and viscous sub steps, satisfying 
\begin{equation}
{D \omega \over Dt}=0 
\label{sub_i}
\end{equation}
and 
\begin{equation}
{\partial \omega \over \partial t}=\frac{1}{Re} \nabla^2 \omega
\label{sub_v}
\end{equation}
respectively.  

The equation for the inviscid sub step represents the fact that for 
two-dimensional inviscid flow vorticity is convected by the flow.  
Numerically, the element vortices are moved at the local fluid velocity, 
i.e.\ 
\begin{equation}
{d {\mathbf z}_j \over dt}~=~ {\mathbf u}({\mathbf x}_j,t)
\label{sub_in}
\end{equation}
A second order Runge-Kutta 
method is used to move the 
vortices at each time step  
\begin{equation}
\hat{\mathbf z}_j ~=~ {\mathbf z}^n_j ~+~\frac{1}{2}\Delta t \,{\mathbf u}({\mathbf z}_j,t_n) 
\label{RK_step1}
\end{equation}
\begin{equation}
{\mathbf z}^{n+1}_j ~=~ {\mathbf z}^n_j ~+~\Delta t \,{\mathbf u}(\hat{\mathbf z}_j,t_{n+1/2}) 
\label{RK_step2}
\end{equation}

A number of methods exist for the viscous sub step (\ref{sub_v}).  
The method developed here is based on the vorticity redistribution 
scheme of Shankar and van Dommelen 
\cite{SvD}.  In this method, at each time the circulation of vortex elements are 
updated through
\begin{equation}
\Gamma_j^{n+1}=\sum_k \Gamma_k^n W^n_{kj}
\label{VR_1}
\end{equation}
where $W^n_{kj}$ represents the fraction of the circulation of vortex 
$k$ transferred to vortex $j$ by diffusion during time step $n$.  
The summation is over a group of vortices local to $z_j$, the position 
of vortex $j$.  The fractions $W^n_{kj}$ are calculated to satisfy 
the following constraints
\begin{equation}
\sum_k\,W^n_{kj}~=~ 1 \label{VR_2}
\end{equation}
\begin{equation}
\sum_k\,W^n_{kj}(x_j-x_k) ~=~ \sum_k \, W^n_{kj}(y_j-y_k) ~=~ 0
\label{VR_3}
\end{equation}
\begin{equation}
\sum_k\,W^n_{kj}(x_j-x_k)^2 ~=~ \sum_k \, W^n_{kj}(y_j-y_k)^2 ~=~ 
2\,  h_v^2 \label{VR_4}
\end{equation}
\begin{equation}
\sum_k\,W^n_{kj}(x_j-x_k)(y_j-y_k) ~=~ 0 
\label{VR_5}
\end{equation}
where $h_v=\surd \overline{\Delta t/Re}$ is the characteristic 
diffusion distance over time $\Delta t$.
Stability requires $W^n_{kj}>0$.  Equations 
(\ref{VR_2}-\ref{VR_5}) enforce conservation of vorticity, 
the centre of vorticity, and linear and angular momentum.  

The redistribution is performed over all vortices within a 
distance  $R h_v$ of vortex $k$, i.e. such that 
\begin{equation}
| z_j-z_k | \, \leq \, R\, h_v
\label{VR_6}
\end{equation}
The accuracy of the method depends on the value of $R$ and 
a minimum value of $R=2$ is required for a first or second 
order solution to exist \cite{SvD}.   
$R=\surd \overline{12}$ was used in \cite{SvD}.  A solution 
may not always exist, for example if there are less than six 
vortices within the region (\ref{VR_6}).  If no solution is 
found, new vortices are introduced a distance 
$\surd\overline{6} h_v$ from the centre ($z_j$) until a solution 
exists.  Further details of the method and its theoretical basis 
can be found in \cite{SvD}.  
 
In this scheme, all vortex elements satisfying (\ref{VR_6}) must 
be identified.  The redistribution problem can then 
be solved using a linear programming 
method, for example the revised simplex scheme found in \cite{simplex}.

\vspace{0.5cm}
\section{The Circulation Redistribution scheme}

\subsection{Redistribution in Cartesian coordinates.}

The redistribution scheme above (\ref{VR_1}-\ref{VR_5}) operates 
by transferring circulation between vortex elements, introducing 
new elements as required.  However, an alternative is to transfer 
the circulation onto a set of nodes at known positions.  
Consider a one-dimensional unsteady 
diffusion problem, with a uniform grid with grid step $h$.  
Suppose there is a vortex 
of strength $\Gamma$ placed at $x=x_v$ where $x_i\leq x_v \leq x_{i+1}$, 
and $x_i=i\,h$.  Let 
\begin{equation}
\Delta ~=~ (x_v-x_i)/h
\label{RD1}
\end{equation}
Then a solution of the redistribution equations is
\begin{equation}
f_i~=~1\,-\,2\left(\frac{h_v}{h}\right)^2 \,-\, \Delta^2
\label{RD2}
\end{equation}
\begin{equation}
f_{i-1}~=~\frac{1}{2}(1\,-\,f_i\, - \,\Delta)
\label{RD3}
\end{equation}
\begin{equation}
f_{i+1}~=\frac{1}{2}(1\,-\,f_i\, + \,\Delta)
\label{RD4}
\end{equation}
where circulation $\Gamma\,f_k$ is transferred to the grid point 
$x=x_k$.  All other $f_k$ are zero.  A second solution is given by 
\begin{equation}
g_{i+1}~=~1\,-\,2\left(\frac{h_v}{h}\right)^2 \,-\, \Delta_1^2
\label{RD5}
\end{equation}
\begin{equation}
g_{i}~=~\frac{1}{2}(1\,-\,g_{i+1}\, - \,\Delta_1)
\label{RD6}
\end{equation}
\begin{equation}
g_{i+2}~=\frac{1}{2}(1\,-\,g_{i+1}\, + \,\Delta_1)
\label{RD7}
\end{equation}
with $g_k=0$ for the other $g_k$, and  
$\Delta_1=(x_v-x_{i+1})/h=\Delta-1$.

In principle either of these solutions could be used.  However, 
consider the case when the element is midway between grid points, 
i.e.\ $\Delta=\frac{1}{2}$.  On physical grounds, 
the redistribution would be expected to be symmetric, but both of the solutions 
are asymmetric.  A simple way of producing a symmetric solution 
in this case is to use the average of the two solutions, i.e.\
$F_k=\frac{1}{2}(f_k\,+\,g_k)$, $k=i-1,\dots,i+2$.  
This gives the same solution as 
would be obtained by using the four points $i-1$ to $i+2$ and 
assuming symmetry ($F_{i-1}=F_{i+2}$ and $F_i=F_{i+1}$).  

More generally, a linear combination of the two basic solutions, 
(\ref{RD2}-\ref{RD4}) and (\ref{RD5}-\ref{RD7}), can be used 
\begin{equation} 
F_k ~=~ (1-\Delta)f_k~+~\Delta g_k,~~~k=i-1,\dots,i+2
\label{RD8}
\end{equation}
This produces a symmetric three point solution if the element is 
at a grid point, and a symmetric four point solution if the 
element is midway between grid points, with a smooth change between 
these two extreme cases.  It is the simplest solution which satisfies 
both the redistribution equations and symmetry.

Some test calculations have been performed for a circular cylinder case 
using using only three point formula ((\ref{RD2}-\ref{RD4}) for 
$x_k\leq x_0 < \frac{1}{2}(x_k+x_{k+1})$ and (\ref{RD5}-\ref{RD7}) 
for $\frac{1}{2}(x_k+x_{k+1}) \leq x_0 < x_{k+1}$), but these produced 
undesirable short scale variations in the solution when a vortex element 
moved across a midpoint and the redistribution changed bias.  This did 
not occur when using the combination (\ref{RD8}).

Stability requires that the redistribution fractions are positive.  The 
most restrictive case when using (\ref{RD8}) is when the element is at a grid 
point.  The condition is then
\begin{equation}
\frac{h_v}{h}\, < \, \frac{1}{\surd 2}
\label{stab1}
\end{equation}

\begin{figure}[h]
 \begin{center}
\includegraphics[scale=0.4]{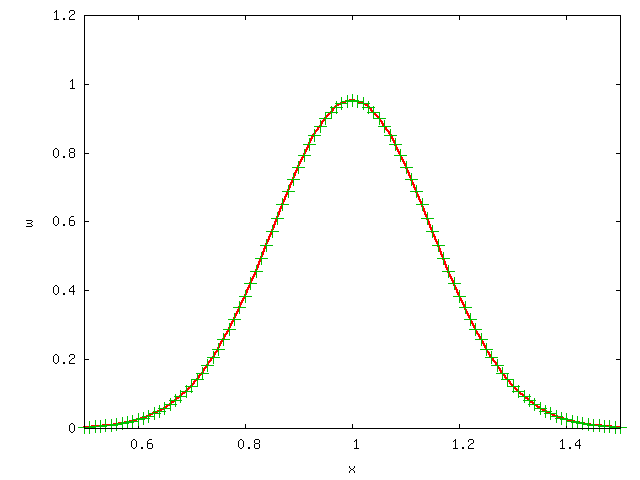}
\caption{Vorticity distribution for the one dimensional test problem 
at $t=1.1$ with $Re=100$, $h=1/100$ and $\Delta t=0.005$.  
The solid line is the analytical solution and the symbols are 
from the redistribution scheme.    }
\label{1D_test}
 \end{center}
\end{figure}

As simple test problem is that of one dimensional diffusion starting from 
a point distribution of vorticity at $x=x_0$ at $t=0$.  Figure 
\ref{1D_test} shows the analytic and redistribution solutions for the 
vorticity for a test case at $t=1.1$ with $x_0=1$, $Re=100$ and total circulation 
of $\surd(Re/\pi)$.  The calculation was started from the analytic solution 
at $t=0.1$.  The grid step was $h=1/100$ and the time step was $t=0.005$.  
Two different grids were used in the calculation.  There was an offset between 
the grids and they were used alternatively at successive time steps.  A number 
of different offsets, ranging from  $\Delta=0$ to $\Delta=1$) were tested, and the 
good agreement with the analytic solution shown in Figure \ref{1D_test} is typical.

Consider now the two dimensional case in $(x,y)$ with grid $x_i=i\,h$ 
and $y_j=j\,\hat{h}$ and a vortex element located at $(x_v,y_v)$ where 
$x_i\leq x_v < x_{i+1}$ and $y_j \leq y_v < y_{j+1}$.  The one dimensional 
redistribution in the $y$ direction is given by
\begin{equation} 
G_l ~=~ (1-\delta)\,\hat{f}_l~+~\delta\, \hat{g}_l,~~~l=j-1,\dots,j+2
\label{RD9}
\end{equation}
where 
\begin{equation}
\delta ~=~ (y_v\,-\, y_j)/\hat{h}
\label{RD10}
\end{equation}
and the $\hat{f}_l$ and $\hat{g}_l$ are obtained as in (\ref{RD2}-\ref{RD7}).

The two dimensional redistribution scheme is given by
\begin{equation}
W_{k,l} ~=~ F_k\,G_l,~~~k=i-1,\dots,i+2,~~l=j-1.\dots,j+2
\label{RD11}
\end{equation}
These weights satisfy all of the constraints in the original redistribution 
scheme (\ref{VR_2}-\ref{VR_5}).

This solution of the redistribution problem satisfies  
Shankar and van Dommelen condition for the existence of 
a solution; the simplest two-dimensional solution  has a nine point stencil  
and the stability condition implies that the corner points 
of the stencil must be a distance of at least $2h_v$ from the 
centre point.  

\begin{figure}[h]
 \begin{center}
\includegraphics[scale=0.4]{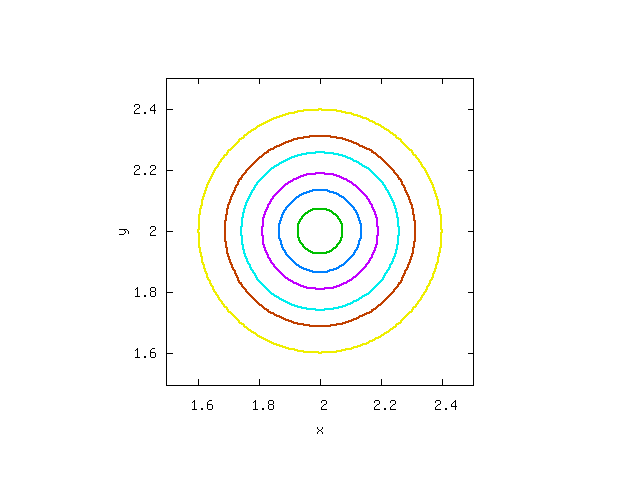}
\caption{Vorticity distribution for the two dimensional test problem 
at $t=1.1$ with $Re=100$, $h=\hat{h}=1/100$ and $\Delta t=0.005$.  
From the centre, the contours are for 0.8, 0.6, 0.4, 0.2, 0.1, and 0.025.    }
\label{2D_test}
 \end{center}
\end{figure}

Figure \ref{2D_test} shows typical constant vorticity contours for the two dimensional 
diffusion problem.  Again, two offset grids were used.  The contours shown are 
from the numerical solution.  At the scale shown they are identical to the 
contours from the analytic solution.

\vspace{0.5cm}
\subsection{Redistribution in cylindrical coordinates}

The standard case of flow past a circular cylinder will be used as a test 
for the method.  The obvious grid to use in this case is one using 
polar coordinates $(r,\theta)$.  However, the original redistribution equations 
must be converted to the appropriate form.  Transforming between polar and Cartesian 
coordinates in standard form $x=r\cos \theta$, $y=r\sin \theta$, write
\begin{equation}
x_k\,=\,(r_j+\Delta r_k)\cos(\theta_j +\Delta \theta_k), ~~~~
y_k\,=\,(r_j+\Delta r_k)\sin(\theta_j +\Delta \theta_k)
\label{polar_1}
\end{equation}

Substituting (\ref{polar_1}) into (\ref{VR_3}-\ref{VR_5}) 
and expanding, while retaining all terms  $O(\Delta^2$) 
or greater, produces the constraints
\begin{equation}
\sum_k\,W^n_{kj}\,r_j\,\Delta r_k~=~ h_v^2 
\label{polar_2}
\end{equation}
\begin{equation}
\sum_k\,W^n_{kj}\,r_j\,\Delta \theta_k~=~ 0 
\label{polar_3}
\end{equation}
\begin{equation}
\sum_k\,W^n_{kj}\,\Delta r_k^2~=~  
\sum_k\,W^n_{kj}\,r_j^2\,\Delta\theta_k^2 ~=~ 2\,h_v^2
\label{polar_4}
\end{equation}
\begin{equation}
\sum_k\,W^n_{kj}\,r_j\Delta r_k\,\Delta \theta_k~=~ 0 
\label{polar_5}
\end{equation}

Equations (\ref{polar_2}-\ref{polar_5}) are analogous to the 
original constants (\ref{VR_3}-\ref{VR_5}) with the 
Cartesian lengths replaced by the polar ones ($\Delta r_k$ and 
$r_v\Delta \theta_k$) apart from the linear radial constraint 
(\ref{polar_2}) which has a nonzero right side.  This term 
arises from the $\frac{1}{r}{\partial \over \partial r}$ in 
the polar Laplace operator; setting the right side of (\ref{polar_2}) 
to zero would produce a nonphysical result in which the diffusion 
operator is  
${\partial^2  \over \partial r^2}+
\frac{1}{r^2} {\partial^2  \over \partial \theta^2}$.

The redistribution formula for the radial one dimensional problem 
are
\begin{equation}
f_i ~=~ 1 \, - \, 2\left(\frac{h_v}{\Delta r}\right)^2 
\, - \, \Delta^2 -2\Delta \frac{h_v^2}{r_v\Delta r}
\label{polar_6}
\end{equation}
\begin{equation}
f_{i-1} ~=~ \frac{1}{2}(1\,-\,f_i\,+\,\Delta\,+\, \frac{h_v^2}{r_v\Delta r} )
\label{polar_7}
\end{equation}
\begin{equation}
f_{i+1} ~=~ \frac{1}{2}(1\,-\,f_i\,-\,\Delta\,-\, \frac{h_v^2}{r_v\Delta r} )
\label{polar_8}
\end{equation}
where the radial grid is given by $r_i-r_0+i\Delta r$, $\Delta=(r_v-r_i)/\Delta R$, 
and the vortex element is at $r=r_v$ where $r_i\leq r_v < r_{i+1}$.

The second redistribution solution involving the points $i$, $i+1$ and $i+2$ 
follows in the obvious manner.  The azimuthal redistribution solution 
is obtained from (\ref{RD2}-\ref{RD6}) with the lengths replaced 
by $r_v\Delta\theta$.  Again, linear combinations of the redistribution 
solutions are formed, and the two dimensional redistribution scheme 
obtained as in (\ref{RD11}).

\subsection{Redistribution in general coordinates.}

Above the redistribution method has been presented for Cartesian 
and polar coordinates system.  The method extends to other, more 
general, systems.  For example, for a transformation given by 
$x=f(\eta)$, the condition for first moment in $x$ becomes 
\begin{equation}
\sum_k\,W^n_{kj}\,\Delta \eta_j\,f^\prime(\eta_k)~=~ -
h_v^2\frac{f^{\prime \prime}(\eta_k)}{f^{\prime 2}(\eta_k)} 
\label{stretch_1}
\end{equation}
where $\Delta \eta_j=\eta_j-\eta_k$ and the other equations are obtained by 
replacing $x_j-x_k$ by $\Delta\eta_jf^\prime(\eta_k)$.   Further, the redistribution 
problem can formulated for a general grid with fixed nodes by calculating 
the appropriate terms, a procedure equivalent to calculating the metric 
terms scaling the derivatives in the diffusion operator in a finite volume 
method.

\vspace{0.5cm}
\section{Computational Algorithm.}

The full scheme is a fractional step algorithm consisting of  
the following steps -- 
\begin{enumerate}
\item
Redistribution onto a fixed grid over a time step of $\frac{1}{2}\Delta t$.
\item
Convection of the vortex elements using the two-step Runge-Kutta scheme 
(\ref{RK_step1},\ref{RK_step2}).
\item
Redistribution onto a fixed grid over a time step of $\frac{1}{2}\Delta t$
\end{enumerate}
 
A method for diffusing vorticity from the surface where it is created 
must be incorporated into this scheme.  A number of methods can 
be found in the literature.  The simplest is to create new vortex 
elements a small distance off the boundary, as in \cite{Spalart,CT}.  
In contrast, \cite{KL} solve a diffusion problem using the 
flux of vorticity from the surface as a boundary condition.  
In \cite{LG}, the redistribution method is used to diffuse 
vorticity from the vortex sheet on the boundary into the interior.  
The last approach is the one used here, in the simplest manner 
consistent with the algorithm.  The circulation from the vortex panels 
created at the start of step 2 is placed on a set of points on the 
boundary (the control points used to calculated the strengths of 
the vortex panels), and these are used as new vortex elements added 
to the redistribution of step 3.  All circulation distributed 
across the boundary  during the redistribution steps is reflected back 
across the boundary, ensuring conservation of circulation and 
imposing a no flux boundary condition.

\vspace{0.5cm}
\section{Body Forces}

The standard way to calculate the lift and drag with a DVM is to use the 
impulse (see e.g.\ \cite{Wu}) -
\begin{equation}
({\cal D,L}) \, = \, -{d \over dt} \int (y\omega, -x\omega) dA 
\, = \, -{d \over dt} \sum_i \Gamma_i(y_i,-x_i)
\label{imp_force}
\end{equation}
where $\cal{D}$ and $\cal{L}$ are the drag and lift normalised by $\rho U_0^2 L$ 
where $L$ is a reference  length.  The summation is performed 
over the entire domain $A$ for all circulation carrying elements.  

The surface pressure can be related to the strength of the vortex panels 
\cite{SS,Spalart} through
\begin{equation}
\Delta p ~=~ \frac{\Delta \Gamma}{\Delta t}
\label{press_force}
\end{equation}
where $\Delta \Gamma$ is the circulation carried by the relevant 
portion of the wall. 

The wall shear stress is obtained by using a finite difference formula 
with velocity components evaluated at fixed points near the 
surface.  For the circular cylinder, the redistribution grid has 
$r_i=r_0+(i-\frac{1}{2})\Delta r$ where $r_0$ is the radius of 
the cylinder.  The shear is calculated using the velocity 
at midpoints through the one sided, second order formula
\begin{equation}
{\partial u_\theta \over \partial r} ~=~ 
{-\,3\,u_\theta(r_0)\,+4\,u_\theta(r_0+\Delta r)\,-\,u_\theta(r_0+2\Delta r) 
\over 2\,\Delta r }
\label{ss_force}
\end{equation}
where $u_\theta$ is the azimuthal velocity.

\vspace{0.5cm}
\section{Test Data: Flow Past a Circular Cylinder.}

There are a large number of papers which use flow past a circular cylinder 
as a test case.  However, in most of these  
only the lift and drag (coefficients) are presented.  
To provide detailed data for comparison,  a finite difference-spectral (FDS) 
code was used.  The streamfunction-vorticity formulation is used, with 
governing equations the vorticity transport equation   (\ref{NS}),  
and the Poisson equation for the streamfunction $\psi$
\begin{equation}
\omega ~=~ -\,\nabla^2 \psi 
\label{Poisson}
\end{equation}

Fourier modes were used in $\theta$, and second order central difference formula 
in the radial direction.  A one sided backwards difference formula, similar 
to (\ref{ss_force}), was used for the time derivative, except for the 
first time step where a backwards Euler scheme was used.  The code is 
fully implicit, iterating to obtain the solution at each time step.  
The radial grid was stretched to give a fine grid near the cylinder, 
and place the outer boundary of the computational domain a long way 
from the surface.   The non-linear terms were handled in the usual 
pseudo-spectral manner.  

With an impulsive start the boundary layer grows as $t^{1/2}$.  This scaling was used 
for the earlier part of the computation, with 
\begin{equation}
r\,-\,r_0 ~=~ 2\left(\frac{t}{Re}\right)^\frac{1}{2}\, \eta,~~~~~
\psi ~=~ t^\frac{1}{2} \Psi,~~~~~
\omega ~=~ t^{-\frac{1}{2}}\, \Omega~
\label{t_scale_1}
\end{equation}
The calculation was switched to a fixed grid at $t=1$, using the radial 
distribution from (\ref{t_scale_1}) at this time.  

This produces a relatively simple but efficient code in which high 
accuracy can be obtained by using a large number of  Fourier modes and 
radial grid points and a small time step.  Grids of up to  1024 complex Fourier 
modes, 2000 radial points, and a time step of $10^{-6}$ 
were used for the data presented below.   Grid independence was checked 
for all Reynolds numbers. 

On the surface of the cylinder 
\begin{equation}
{\partial p \over \partial \theta } ~=~ {r_0 \over Re} \, 
{\partial \omega \over \partial r}
\label{pg_fd}
\end{equation}
which can be used to calculate the surface pressure 
using a reference value of zero at the front of the cylinder.  
This equation is analogous to (\ref{press_force}) for the DVM.  
Both relate the flux of vorticity from the surface to the 
pressure gradient.  

The results produced by this code compare well with those found 
in other high resolution simulation (e.g.\ \cite{KL}).  Also, 
they agree with the short time series solutions given by 
\cite{BLY,SCRD}.

\vspace{0.5cm}
\section{Choice of grid and numerical parameters.}

As a test case for the effects of the numerical parameters and grid 
on the accuracy of the solution, the drag for the flow past an impulsively 
started cylinder for short time will be used.  Lengths are scaled or the 
diameter of the cylinder $D$ so that the surface of the cylinder is at 
$r=r_0=1/2$, the vorticity is scaled by $U_0/D$ where $U_0$ is the 
free stream velocity, and the time is scaled by $D/U_0$.  The Reynolds number 
is $Re=U_0\,D/\nu$ where $\nu$ is the kinematic viscosity.  
 
A body fitted polar grid is used in the region $r_0\leq r \leq r_1$.  This is 
embedded in a uniform Cartesian mesh for $r>r_1$.  
The inner grid is arranged so that the surface of the cylinder falls midway 
between radial grid points, with $r_i=r_0+(i-\frac{1}{2})\Delta r$ where 
$\Delta r$ is the radial grid step, so that the surface is at $r=r_{1/2}$.  
Azimuthally, the grid is placed at uniformly spaced points $\theta=\theta_j$ 
with grid step $\Delta \theta=2 \pi/N$ where $N$ is the number of vortex 
panels.  The end of the vortex panels are 
at $\theta=\theta_{j+1/2}$, with the control points for the evaluation 
of the boundary velocity at $(r_0,\theta_j)$.  

The method does not explicitly allow for the $t^\frac{1}{2}$ behaviour for small 
time, so the solution cannot be expected to be accurate over the first few 
time steps.  However, with an appropriate choice of parameters, solutions 
which achieve high accuracy after a few  time steps can be obtained.  

There are six numerical 
parameters which must be chosen.  The grid steps in $r$ and $\theta$ for the inner 
grid, the grid step $h$ for the outer grid (a square grid is used here although 
this is not required), the region for the inner grid ($r_1$), 
the time step  $\Delta t$ and the core size $\sigma$.  

The  maximum time step is fixed by the stability of the redistribution scheme.  
Since the algorithm 
has two redistribution substeps over time $\Delta t/2$, 
the stability condition becomes $h_v/h_m<1$, or $\Delta t< Re\, h_m^2$,  
where $h_m$ is the smallest of the three grid steps $\Delta r$, $r_0\Delta \theta$ 
and $h$. 

The effects of the core size were investigated in \cite{CT}, and they concluded 
that taking $\sigma=l/4$ where $l$ is the (average) length of a vortex panel 
was a suitable choice.  This works well here also, giving 
$\sigma=r_0\Delta \theta/4$.

\begin{figure}
 \begin{center}
\includegraphics[scale=0.35]{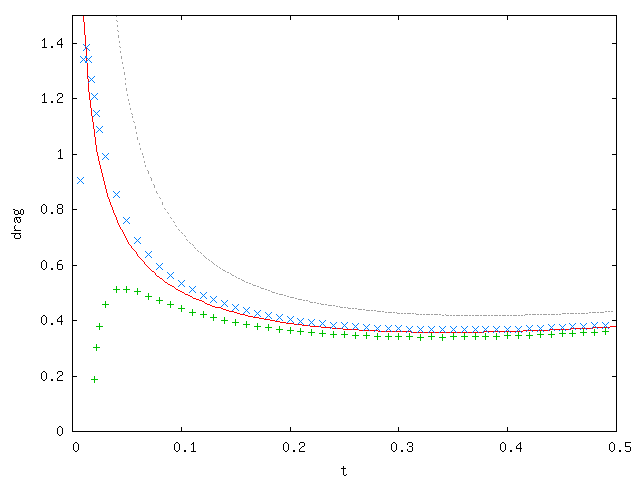}
\caption{The drag for impulsively started 
flow past a circular 
cylinder at $Re=550$  cylinder at $Re=550$ with 400 vortex panels, $r_1=\frac{6}{5}$, 
$\Delta t =0.0025$ and 
$h=\frac{1}{80}$:
dashes, $\Delta r = \frac{1}{200}$;  $\times$, $\Delta r =\frac{1}{300}$; 
$+$, $\Delta r = \frac{1}{400}$.  Solid line, FDS solution.  }
\label{drag_Re550_400}
 \end{center}
\end{figure}

The effect of the grid on the solution was investigated by fixing the number 
of vortex panels and varying $\Delta r$.  Figure \ref{drag_Re550_400} 
shows the drag calculated from the impulse (\ref{imp_force}) 
for short time for an impulsive start with 
$Re=550$, 400 panels, $\Delta t =0.025$, and $\Delta r=1/200$, $1/300$ and 
$1/400$.  For the smallest value of $\Delta r$, $h_v/\Delta r \approx 0.85$.  
Apart from the first few steps, the middle value ($\Delta r = 1/300$) gives 
good agreement from the drag for the FDS scheme, while for the smaller value  
($\Delta=1/400$) the drag approaches that from  the FDS scheme from below.  
For $\Delta = 1/200$, the drag is too large.   For clarity, 
only every fourth point is shown for the solutions for $h=1/300$ and $1/400$.  
All points are used for $h=1/200$, and no filtering or smoothing has 
been applied for this figure.  The smoothness is typical of the results 
obtained when using a body fitted redistribution mesh.

\begin{figure}
 \begin{center}
\includegraphics[scale=0.35]{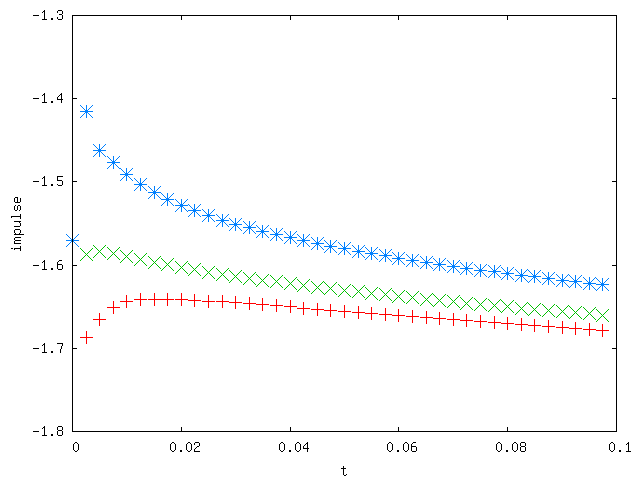}
\caption{The streamwise component of impulse for impulsively started 
flow past a circular 
cylinder at $Re=550$ with 400 vortex panels, $\Delta t =0.0025$ and 
$h=\frac{1}{80}$:
$*$, $\Delta r = \frac{1}{200}$;  $\times$, $\Delta r =\frac{1}{300}$; 
$+$, $\Delta r = \frac{1}{400}$    }
\label{impulse_Re550_400}
 \end{center}
\end{figure}

Figure \ref{impulse_Re550_400} shows the streamwise component of the impulse, 
$I_x=\int y\omega dA$, for the same cases as in Figure \ref{drag_Re550_400}.  
For an impulsive start for flow past a cylinder, the initial condition  
at $t=0^+$ is that from potential flow, with a vortex sheet of (nondimensional) strength 
$\Gamma=-2\sin(\theta)$ on the surface, giving $I_x=-\pi/2$.  
For $t>0$, the impulse should decrease smoothly from the initial value. 
For $\Delta r = 1/300$, there is some (expected) irregular behaviour 
for the first few time steps, but the impulse is generally well behaved.  
In contrast, the smaller and larger values of $\Delta r$ produce a jump 
in the impulse, followed by a relatively fast decrease for $\Delta r = 1/200$ 
and an increase for $\Delta r = 1/400$, consistent with the behaviour of 
the drag (Figure \ref{drag_Re550_400}).

Calculations were performed for with different time steps (smaller for all three 
values of $\Delta r$ and larger for $\Delta r = 1/200$ and $1/300$), but the 
behaviour of the impulse was similar to that shown in Figure \ref{impulse_Re550_400} with 
an overshoot for $\Delta r = 1/200$ and an overshoot for $\Delta r =1/300$.   
The impulse was examined for a large number of other runs  
with Reynolds numbers varying from 150 to 9500 and a range of time steps,   
and its behaviour for short time provides a useful diagnostic 
as to the quality as the grid with regard to the ratio of the grid steps. This 
test could be used for other problems in which there is no reliable 
solution available to compare with, e.g.\ for flow past a square 
(Section \ref{square_results} below).

\begin{figure}
 \begin{center}
\includegraphics[scale=0.35]{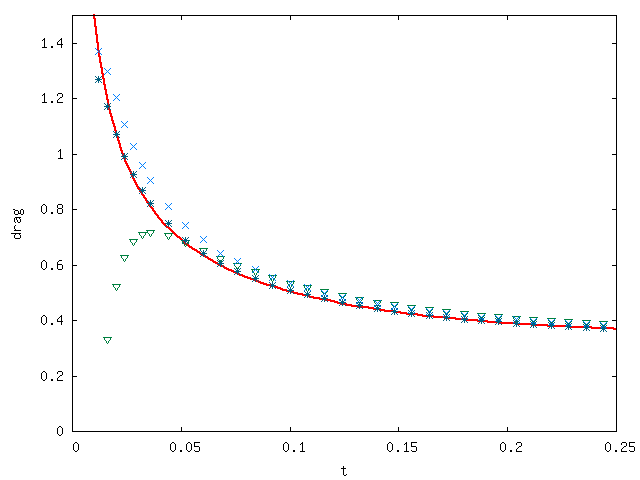}
\caption{The drag for impulsively started 
flow past a circular 
cylinder at $Re=550$  cylinder at $Re=550$ with  
$\Delta t =0.002$; solid line, FDS solution; 
$\triangle$, 200 panels; $\times$, 400 panels; $*$, 600 panels. }
\label{drag_Re550_grid}
 \end{center}
\end{figure}

For all the Reynolds numbers studied, a ratio of approximately 2/5 for the radial to azimuthal 
grid ($\Delta r/r_0\Delta \theta$) with $\sigma=r_0\Delta \theta/4$ was found to  
give an accurate solution provided the time step was  
small enough.  Figure \ref{drag_Re550_grid} shows the drag for runs with 
$\Delta t = 0.002$, 200 panels and $\Delta r = 1/150$, 400 panels and 
$\Delta r = 1/300$, and 600 panels and $\Delta r = 1/450$, i.e.\  maintaining the 
same scaling as for 400 panels in Figures~\ref{drag_Re550_400} and  
\ref{impulse_Re550_400}.  Clearly, 
the grid is too coarse to provide a good match with the FDS solution 
very early in the run, but does give a reasonable value for $t>0.1$.  
As above, there is good match with 400 panels, and a very close match with 600 panels.  

The outer grid step $h$ and $r_1$ were also varied to ensure they did 
not significantly affect the results shown in 
Figures \ref{drag_Re550_400}-\ref{drag_Re550_grid}.   
Tests were also performed with other Reynolds numbers to ensure 
the results presented below are accurate.

\vspace{0.5cm}
\section{DVM solutions: Flow Past a Circular Cylinder.} \label{cylinder_results}

Calculations were performed for impulsively started flow past a circular cylinder 
using Reynolds numbers of 150, 550, 1000, 3000, and 9500.  These Reynolds numbers 
were chosen as they are commonly used as test cases.  
A large of amount of test data was generated, showing excellent agreement 
with the results from the FDS code in all cases (and with data found in other 
studies). 
Representative results are presented for three Reynolds numbers 
($Re=150$, 1000, and 9500), covering three orders of magnitude.

\subsection{$\mathbf{Re=150}.$}

\begin{figure}
 \begin{center}
\includegraphics[scale=0.35]{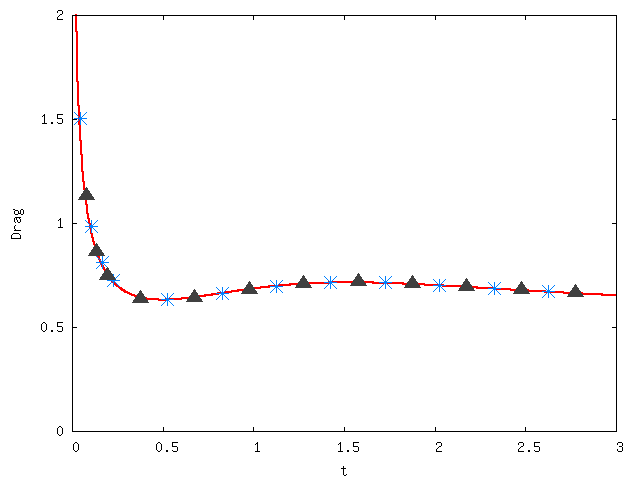}
\caption{Total drag for impulsively started flow past a circular 
cylinder at $Re=150$.  Line: FDS solution.  Symbols: DVM solution, $*$ 
from the impulse(\ref{imp_force}), $\triangle$ from the surface forces 
(\ref{press_force}-\ref{ss_force}).     }
\label{total_drag_Re150}
 \end{center}
\end{figure}

\begin{figure}
 \begin{center}
\includegraphics[scale=0.35]{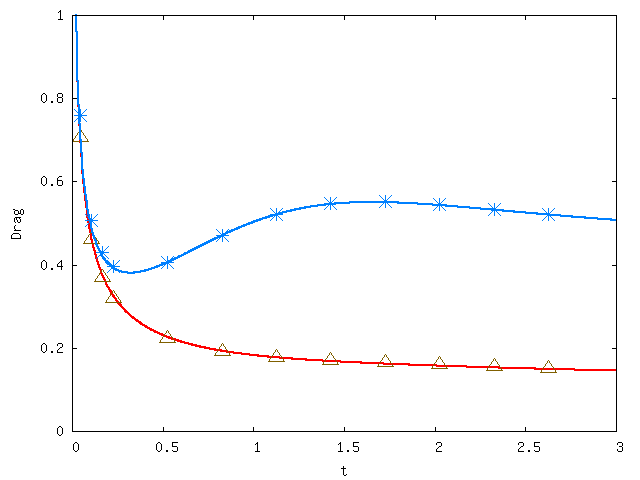}
\caption{Drag components for impulsively started flow past a circular 
cylinder at $Re=150$.   Upper, pressure component.  Lower, shear stress component. 
Lines: FDS solution.  Symbols: DVM solution.   }
\label{drag_components_Re150}
 \end{center}
\end{figure}

Figure \ref{total_drag_Re150} shows the total drag obtained from the 
DVM and FDS methods for flow with $Re=150$.  The numerical parameters are 
$N=400$, $\Delta r = 1/320$, $\Delta t = 0.001$ ($h_v/\Delta r =0.32$), 
$r_1=1$ and $h=1/100$.  
For the DVM, both the 
drag from the impulse 
(\ref{imp_force}) and that from the pressure and wall shear stress 
(\ref{press_force}-\ref{ss_force}) are shown.  There is excellent agreement 
between all methods.    Figure \ref{drag_components_Re150} shows the 
pressure and wall shear stress components of the drag obtained from the DVM and 
FDS schemes.  Again there is excellent agreement.  

\begin{figure}
 \begin{center}
\includegraphics[scale=0.35]{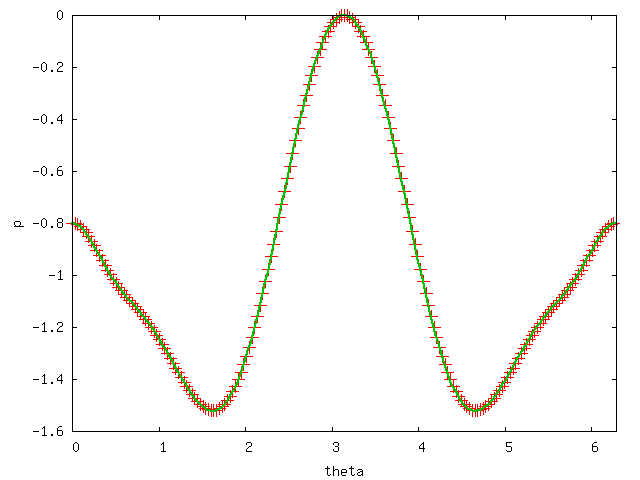}
\caption{Surface pressure $p$ against $\theta$ at $t=1$ for $Re=150$.  
$\theta$ is in radians 
measured from the rear of the cylinder.  
The reference value 
is zero at the front of the cylinder ($\theta=\pi$).    
Symbols: DVM solution, line: finite difference solution. }
\label{pressure_Re150}
 \end{center}
\end{figure}

\begin{figure}
 \begin{center}
\includegraphics[scale=0.35]{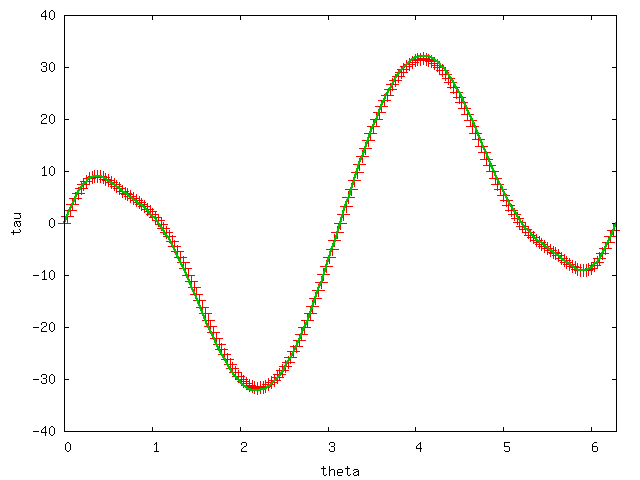}
\caption{Wall shear stress against $\theta$ at $t=1$ for $Re=150$.    
Symbols: DVM solution, line: finite difference solution. }
\label{tau_Re150}
 \end{center}
\end{figure}

\begin{figure}
 \begin{center}
\includegraphics[scale=0.5]{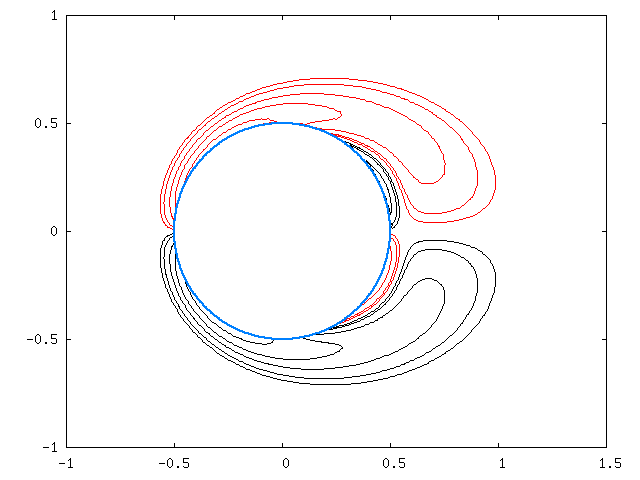}
\caption{Vorticity contours for $Re=150$ at $t=1$: black, positive;
red, negative.  Going from the far field towards the 
cylinder, the contours are for $|\omega |$=1,2,5,10,20.   
The contours above the axis ($y>0$) are from the DVM code, 
and those below the axis from the FDS code. }
\label{cont_both_Re150_t1}
 \end{center}
\end{figure}

The surface distribution of pressure and wall shear stress at 
a single point in time ($t=1$) are shown in Figures \ref{pressure_Re150} 
and \ref{tau_Re150}.  There is very good agreement between the values 
from the two numerical schemes. 
Figure \ref{cont_both_Re150_t1} shows contours of the vorticity at 
$t=1$, with the upper half of the plot showing the contours from the 
DVM method and the lower from the FDS scheme.  Again, there is excellent 
agreement.

\subsection{$\mathbf{Re=1000}.$}

\begin{figure}
 \begin{center}
\includegraphics[scale=0.35]{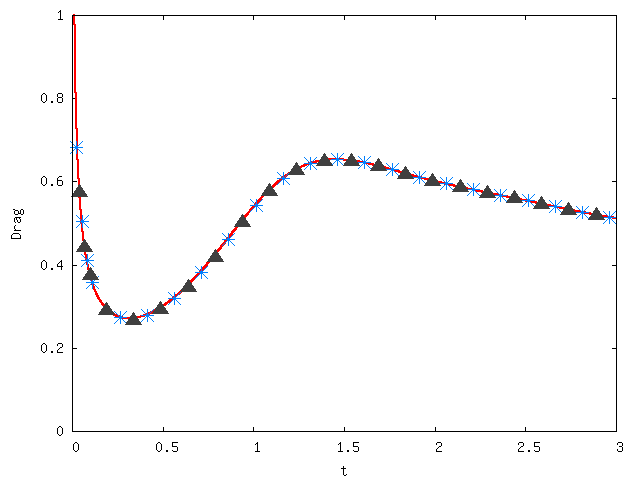}
\caption{Total drag for impulsively started flow past a circular 
cylinder at $Re=10^3$.  Line: FDS solution.  Symbols: DVM solution, $*$ 
from the impulse(\ref{imp_force}), $\triangle$ from the surface forces 
(\ref{press_force}-\ref{ss_force}).  }
\label{total_drag_Re1000}
 \end{center}
\end{figure}

\begin{figure}
 \begin{center}
\includegraphics[scale=0.35]{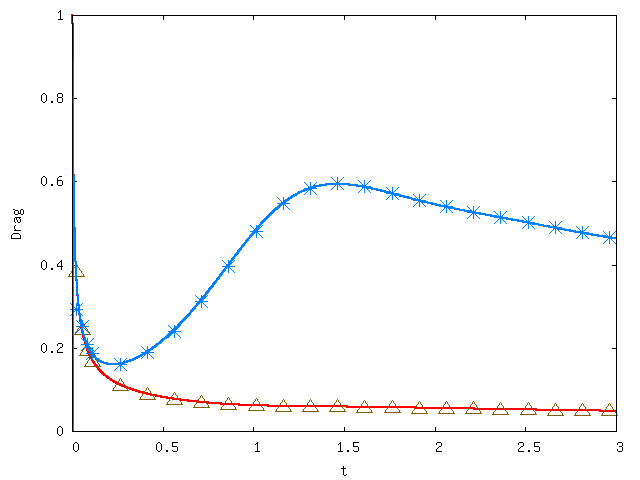}
\caption{Drag components for impulsively started flow past a circular 
cylinder at $Re=10^3$.   Upper, pressure component.  Lower, shear stress component. 
Symbols: DVM solution. Lines: finite difference solution.   }
\label{drag_components_Re1000}
 \end{center}
\end{figure}

\begin{figure}
 \begin{center}
\includegraphics[scale=0.35]{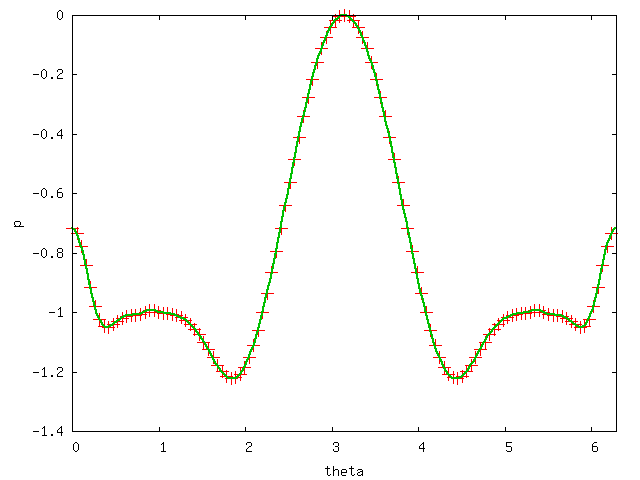}
\caption{Surface pressure $p$ against $\theta$ at $t=3$ for $Re=10^3$.  
$\theta$ is in radians 
measured from the rear of the cylinder.  
The reference value 
is zero at the front of the cylinder ($\theta=\pi$).    
Symbols: DVM solution, line: finite difference solution. }
\label{pressure_Re1000}
 \end{center}
\end{figure}

\begin{figure}
 \begin{center}
\includegraphics[scale=0.35]{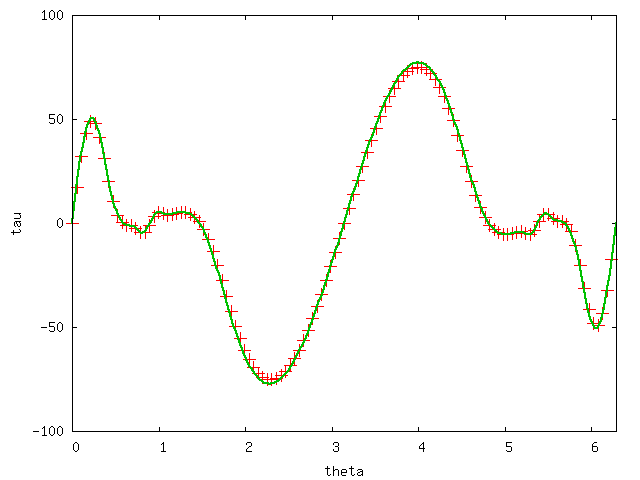}
\caption{Skin friction against $\theta$ at $t=3$ for $Re=10^3$.    
Symbols: DVM solution, line: finite difference solution. }
\label{tau_Re1000}
 \end{center}
\end{figure}

\begin{figure}
 \begin{center}
\includegraphics[scale=0.5]{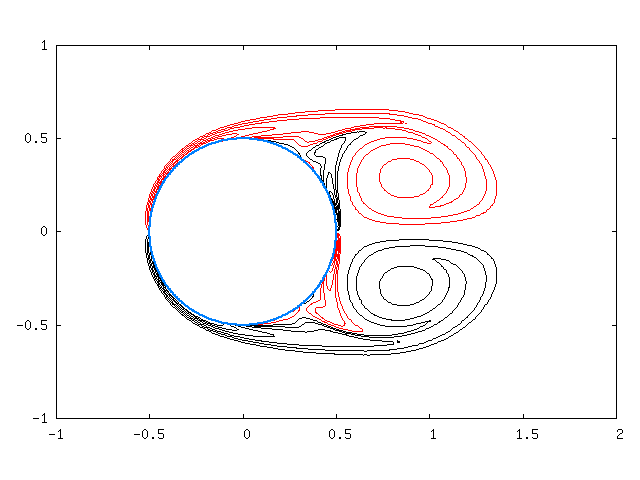}
\caption{Vorticity contours for $Re=10^3$ at $t=3$: black, positive;
red, negative.  Going from the far field towards the 
cylinder, the contours are for $|\omega |$=2,5,10,30,40. 
The contours above the axis ($y>0$) are from the DVM code, 
and those below the axis from the FDS code. }
\label{cont_both_Re1000_t3}
 \end{center}
\end{figure}

Figures \ref{total_drag_Re1000} and \ref{drag_components_Re1000} show the total 
drag and drag components for the two methods for flow with $Re=1000$.  
The numerical parameters are $N=600$,  $\Delta r = 1/480$, 
$\Delta t = 0.0025$ ($h_v/\Delta r \approx 0.76$), 
$r_1=1$ and $h=1/480$. 
As for $Re=150$, there is excellent agreement.  Also, very good 
agreement is obtained for the surface forces 
and contours of vorticity, as can be seen 
for $t=3$ in Figures \ref{pressure_Re1000}, \ref{tau_Re1000} and 
\ref{cont_both_Re1000_t3}.

\subsection{$\mathbf{Re=9500}.$}

Flow with $Re=9500$ provides a much stiffer test of the method as the 
flow field is more complex, with short scale variations in the surface forces 
and a much more complicated vorticity pattern than that found with lower Reynolds numbers. 
Again, however, there is very good agreement between the solutions for the two 
numerical schemes.

\begin{figure}
 \begin{center}
\includegraphics[scale=0.35]{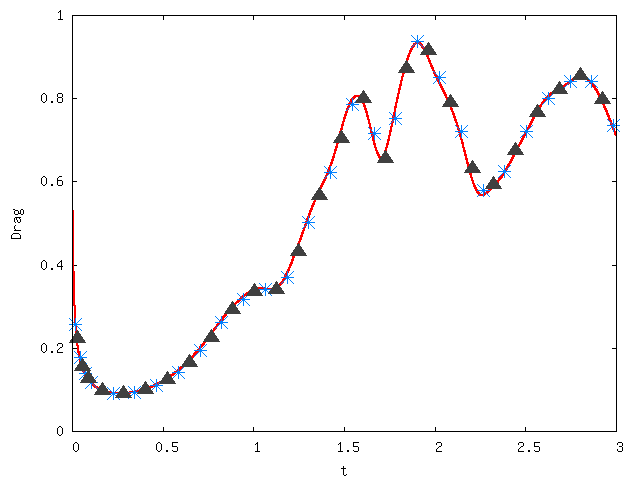}
\caption{Total drag for impulsively started flow past a circular 
cylinder at $Re=9500$.   Line: FDS solution.  Symbols: DVM solution, $*$ 
from the impulse(\ref{imp_force}), $\triangle$ from the surface forces 
(\ref{press_force}-\ref{ss_force}).  }
\label{total_drag_Re9500}
 \end{center}
\end{figure}

\begin{figure}
 \begin{center}
\includegraphics[scale=0.35]{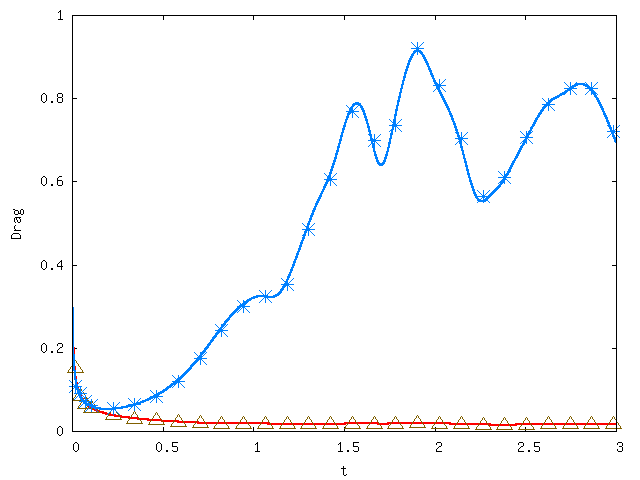}
\caption{Drag components for impulsively started flow past a circular 
cylinder at $Re=9500$.  
Symbols: DVM solution. Lines: finite difference solution.   }
\label{drag_components_Re9500}
 \end{center}
\end{figure}

\begin{figure}
 \begin{center}
\includegraphics[scale=0.35]{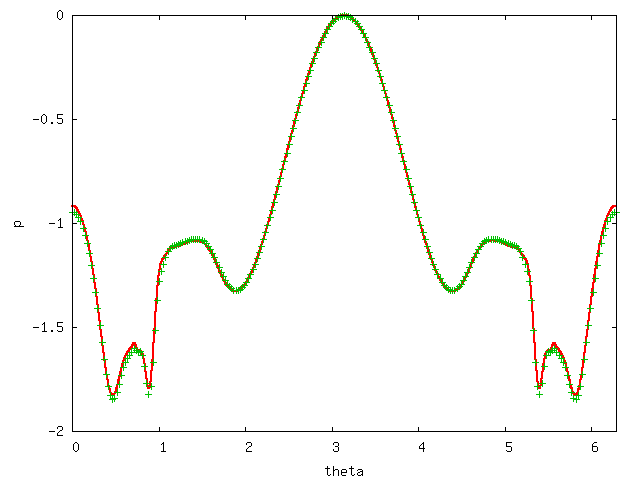}
\caption{Surface pressure $p$ against $\theta$ at $t=2$ for $Re=9500$.  
$\theta$ is in radians 
measured from the rear of the cylinder.  
The reference value 
is zero at the front of the cylinder ($\theta=\pi$).    
Symbols: FDS solution, line: DVM solution. }
\label{pressure_Re9500}
 \end{center}
\end{figure}

\begin{figure}
 \begin{center}
\includegraphics[scale=0.35]{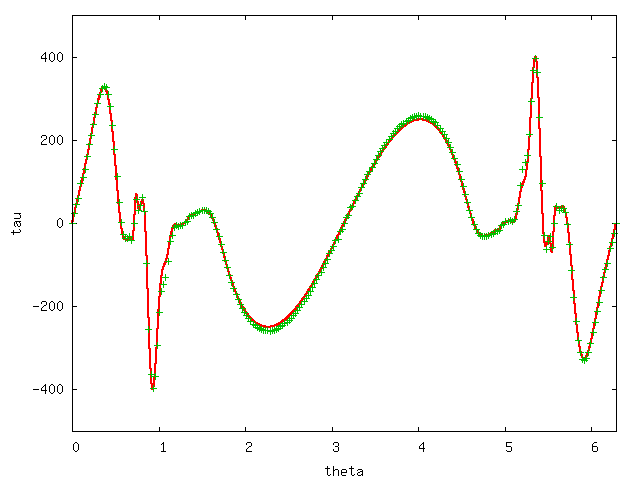}
\caption{Skin friction against $\theta$ at $t=2$ for $Re=9500$.    
Symbols: FDS solution, line: DVM solution. }
\label{tau_Re9500}
 \end{center}
\end{figure}

\begin{figure}
 \begin{center}
\includegraphics[scale=0.45]{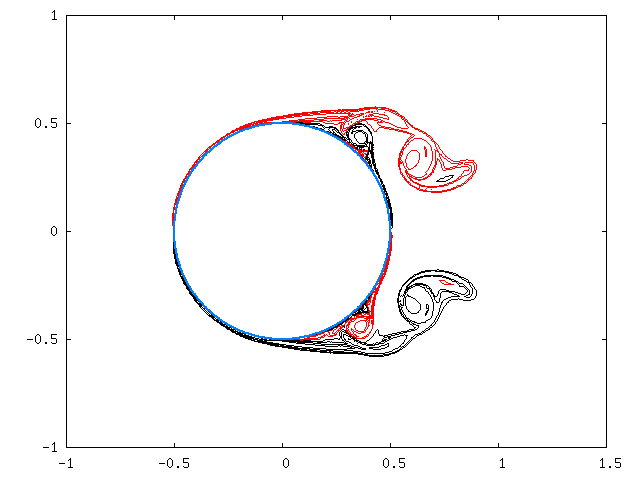}
\caption{Vorticity contours for $Re=9500$ at $t=2$: black, positive;
red, negative.  Going from the far field towards the 
cylinder, the contours are for $|\omega |$=5,10,25,50,100,200.  
The contours above the axis ($y>0$) are from the DVM code, 
and those below the axis from the FDS code.  }
\label{cont_both_Re9500_t2}
 \end{center}
\end{figure}

Figures \ref{total_drag_Re9500} and \ref{drag_components_Re9500} show the total 
drag and drag components. 
The numerical parameters are $N=3000$,  $\Delta r = 1/2400$, 
$\Delta t = 0.001$ ($h_v/\Delta r \approx 0.78$), 
$r_1=6/5$ and $h=1/800$. 
The surface forces at $t=2$ are shown in Figures 
\ref{pressure_Re9500}, \ref{tau_Re9500}.  There is a high level 
of agreement, in particular, in the wall shear stress on the rear part of the cylinder 
where the development of relatively small scale but strong structures in the 
flow lead to large peaks and high values of the gradient along the surface.  
The complex nature of the flow can also be seen in the vorticity contours  
(Figure \ref{cont_both_Re9500_t2}).

The  values of the wall shear stress at the top and bottom shoulder of the 
cylinder ($\theta = \pi/2$ and $3\pi/2$) are slightly lower for the DVM method as 
compared with those for from the FDS calculations.  However, the radial grid 
used for the DVM calculation near the surface is coarse 
as compared to that for FDS, and it was found that increasing the resolution 
gave a better match, but with an increase in computational effort.

This run generated  
approximately $8.4\times10^5$ vortex element by $t=3$, a similar number to that 
used by Koumoutsakos and Leonard \cite{KL}.  For comparison, 
for $Re=150$ (Figures \ref{total_drag_Re150}-\ref{cont_both_Re150_t1}), 
there were approximately $9.5\times10^4$ vortex elements at 
$t=3$.

\vspace{0.5cm}
\section{Flow past a square.} \label{square_results}

There is relatively little data available for flow past a square as 
compared to that for a cylinder.  However, plots of drag and vorticity contours 
are given in \cite{PW} for an impulsive start with $Re=100$ and the square at 
15$^\circ$ angle of attack. A similar calculation was performed, using a uniform, body 
fitted, Cartesian grid embedded in a uniform Cartesian grid aligned the flow.  
Eighty one constant length vortex panels were used on each side of the body, 
and a grid step of $h=1/81$ was used for both the inner outer grids.  The change 
between the inner and outer grids occurred a distance 0.5 from the body. The time step was 
$\Delta t = 0.005$, giving $h_v/h\approx 0.57$.  With this set of parameters, 
the behaviour of the impulse early in the calculation was as expected.     

\begin{figure}
 \begin{center}
\includegraphics[scale=0.35]{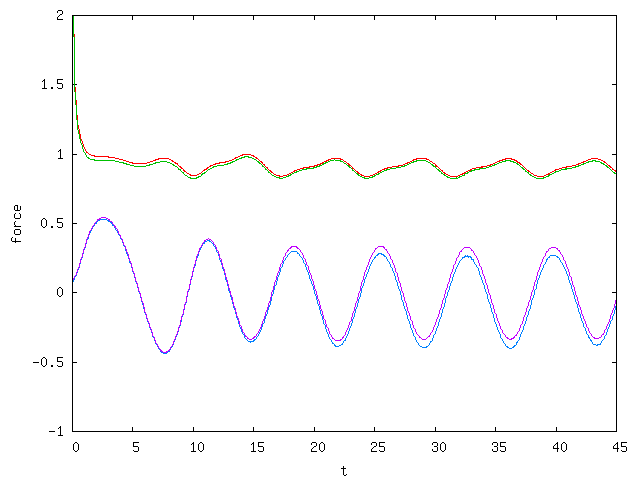}
\caption{Lift and drag for impulsively started flow past a square  
at 15$^\circ$ and $Re=100$.  The top two lines are the drag, with the upper one from 
impulse and the lower from the surface forces.  The bottom two lines are 
the lift, with the lower from the impulse and the upper from the surface forces.  
  }
\label{force_square_Re100}
 \end{center}
\end{figure}

The flow at a right angled convex corner is singular, but both the pressure 
and the vorticity behaving as $\rho^{-0.456}$ where $\rho$ is the distance from the 
corner \cite{Moffat}.  Hence, there may be large errors in the surface pressure 
obtained by integrating the panel strengths, and in calculating the lift 
and drag from the surface forces.  Figure \ref{force_square_Re100} shows the 
drag and lift obtained from both methods.  There is reasonable agreement, given the 
potential for large errors.  The drag is consistent with that given in \cite{PW}.  
A calculation with 41 panels on each side of the square and $h=1/41$ produced 
a similar result to that shown in Figure \ref{force_square_Re100}, 
but with a larger difference between the drag and lift  calculated from the impulse 
and the surface forces.

\begin{figure}
 \begin{center}
\includegraphics[scale=0.5]{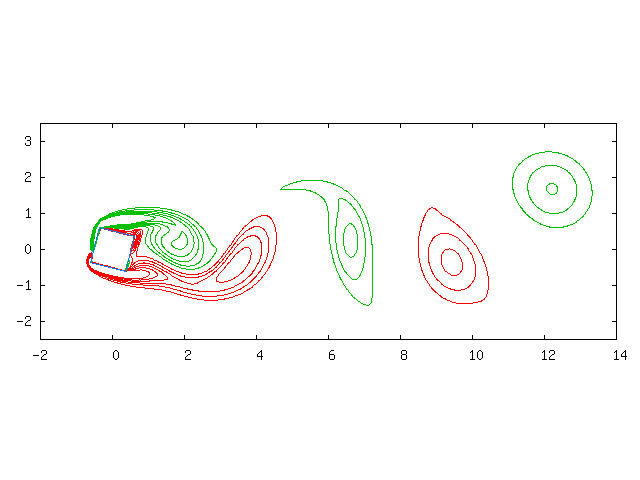}
\caption{Vorticity contours at $t=20$ for impulsively started flow 
past a square at 15$^\circ$ and Re=100.  A step of 0.5 is used 
with zero omitted.   }
\label{cont_square_Re100_t20}
 \end{center}
\end{figure}

Vorticity contours for for $t=20$ are shown in Figure \ref{cont_square_Re100_t20}.  
This figure agrees well with that in \cite{PW}, in particular, as regards the 
position and strength of the vortices downstream of the body.

In \cite{PW}, the corners of the square were rounded to avoid unspecified 
numerical problems.  This was not required for the calculations performed in the 
current work.

\vspace{0.5cm}
\section{DVM scheme with a single grid.}

All of the calculations described above have used a boundary fitted redistribution 
mesh near the body and a regular Cartesian mesh further away.  Care has been taken 
to ensure that the total circulation is conserved.  An alternative approach, used in a 
number of previous studies (e.g.\ \cite{Spalart,CT,PW}), 
is to delete any vorticity that crosses the boundary and rely on the creation 
process to regenerate the vorticity in an appropriate manner.  
There are several advantages to this approach.  In particular, it allows 
simulation for flow past bodies of an arbitrary shape by embedding them into 
a regular grid, and simply deleting any vortex elements which are redistributed into the 
body.  The major disadvantage is that the method will no longer produce 
high resolution values for the surface forces.  In particular, since part of the 
circulation in the vortex sheet arises from the non conservative nature of the 
redistribution at the surface of the body, the surface pressure cannot be 
estimated using (\ref{press_force}) unless the deletion is accounted for.  

Simulations were performed using a circular cylinder embedded in a uniform Cartesian 
grid.  Following \cite{CT}, the vortex elements created each time step were placed a distance 
of $1.12\sigma$ above the surface so that the maximum velocity 
generated by a new element occurs at the surface, while all vortex elements  
within this distance or below the surface were deleted after the redistribution.  

\begin{figure}
 \begin{center}
\includegraphics[scale=0.35]{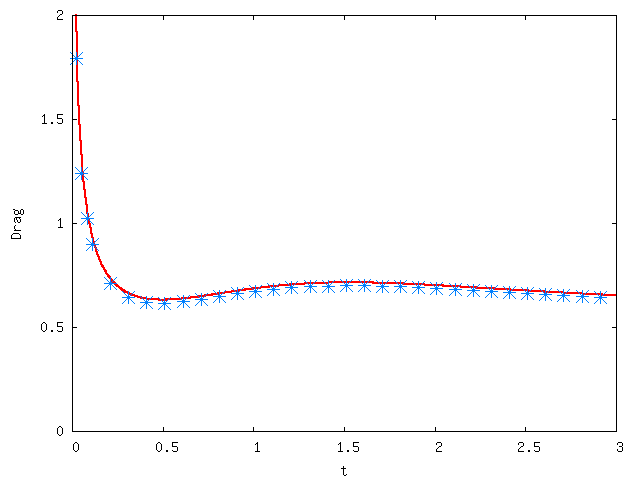}
\caption{Total drag for impulsively started flow past a circular 
cylinder at $Re=150$.  Line: FDS solution.  Symbols: DVM solution using a 
single Cartesian redistribution grid with $h=0.01$, $\Delta t = 0.005$ 
and 400 vortex panels.
}
\label{total_drag_Re150_single}
 \end{center}
\end{figure}

\begin{figure}
 \begin{center}
\includegraphics[scale=0.4]{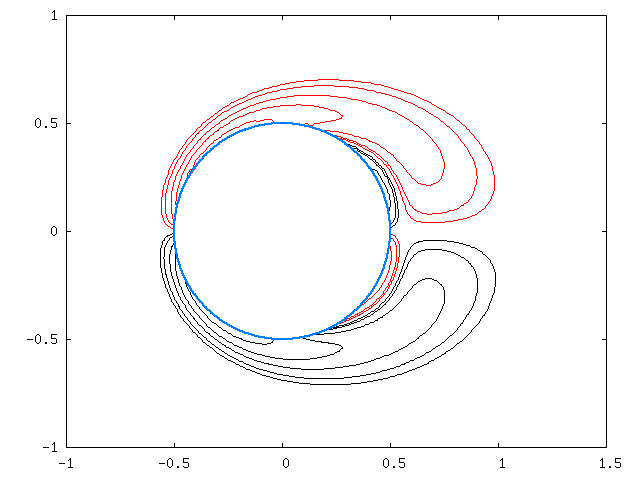}
\caption{Vorticity contours for $Re=150$ at $t=1$: black, positive;
red, negative.  Going from the far field towards the 
cylinder, the contours are for $|\omega |$=1,2,5,10,20.   
The contours above the axis ($y>0$) are from the DVM code with a single 
Cartesian redistribution grid with $h=0.01$, $\Delta t = 0.005$ and 400 vortex panels, 
and those below the axis from the FDS code. }
\label{cont_both_Re150_single_t1}
 \end{center}
\end{figure}

The drag for a flow with $Re=150$, 400 vortex panels, a time step of 
$\Delta t = 0.005$ and a redistribution mesh with $h=0.01$ is shown in 
Figure \ref{total_drag_Re150_single}.  Also shown is the drag from the FDS 
scheme, showing good agreement with the DVM values.  The vorticity 
distribution for both methods at $t=1$ is shown in Figure  
\ref{cont_both_Re150_single_t1}. Overall there is very close agreement, although 
some differences can be seen near the surface.  
However, and as expected, the distribution of vortex panel strengths and the wall 
shear stress showed large high frequency oscillations.      

\begin{figure}
 \begin{center}
\includegraphics[scale=0.4]{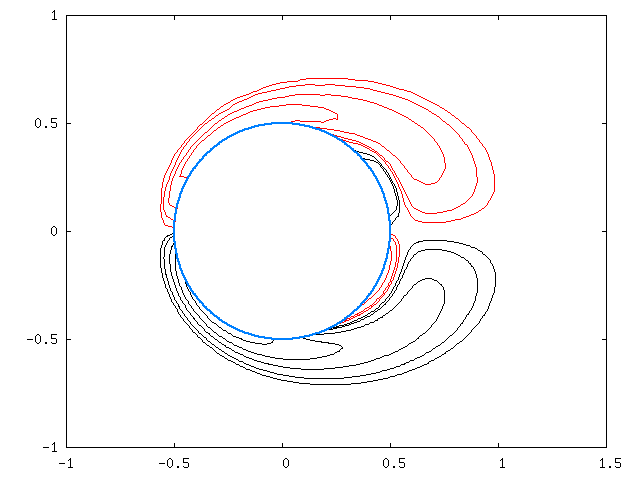}
\caption{Vorticity contours for $Re=150$ at $t=1$: black, positive;
red, negative.  Going from the far field towards the 
cylinder, the contours are for $|\omega |$=1,2,5,10,20.   
The contours above the axis ($y>0$) are from the DVM code with a single 
Cartesian redistribution grid with $h=0.02$, $\Delta t = 0.02$ and 200 vortex panels, 
and those below the axis from the FDS code. }
\label{cont_both_Re150_single_t1_coarse}
 \end{center}
\end{figure}

A further calculation was performed for $Re=150$ but with 200 panels 
and $h=0.02$ so that there were approximately quarter the number of vortex elements.  
The drag was almost the same as shown in Figure \ref{total_drag_Re150_single}.  
The vorticity contours at $t=1$ for both the DVM and FDS schemes are shown 
in Figure \ref{cont_both_Re150_single_t1_coarse}.  Away from the body there is 
still good agreement between the two solutions but the lack of resolution 
near the surface with the DVM method is more apparent.

\begin{figure}
 \begin{center}
\includegraphics[scale=0.35]{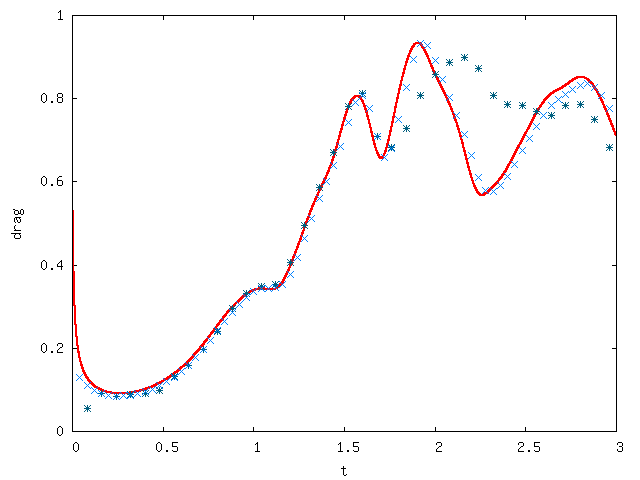}
\caption{Total drag for impulsively started flow past a circular 
cylinder at $Re=9500$.  Line: FDS solution.  Symbols: DVM solutions using a 
single Cartesian redistribution grid: $\times$, 3000 panels, $h=1/1000$ 
and $\Delta t = 0.005$; $*$, 1600 panels, $h=1/400$ and 
$\Delta t = 0.02$. }
\label{drag_Re9500_single}
 \end{center}
\end{figure}

Calculations were also performed for flow with $Re=9500$.  Figure 
\ref{drag_Re9500_single} shows the drag for the FDS method and the DVM 
scheme with two different resolutions.  The better resolved DVM 
calculation has $N=3000$, $h=1/1000$ and $\Delta t = 0.005$, 
and the coarser calculation $N=1600$, $h=1/400$ and $\Delta t =0.02$.  
In both cases, the redistribution grid step was chosen to approximately 
match the vortex panel length. The drag from the better resolved DVM solution 
shows good agreement with that from the FDS method, except, unsurprisingly, 
during the very early part of the run.   The drag from the coarser calculation 
also agrees well up to $t\approx 1.5$, but not at later times.

\begin{figure}
 \begin{center}
\includegraphics[scale=0.4]{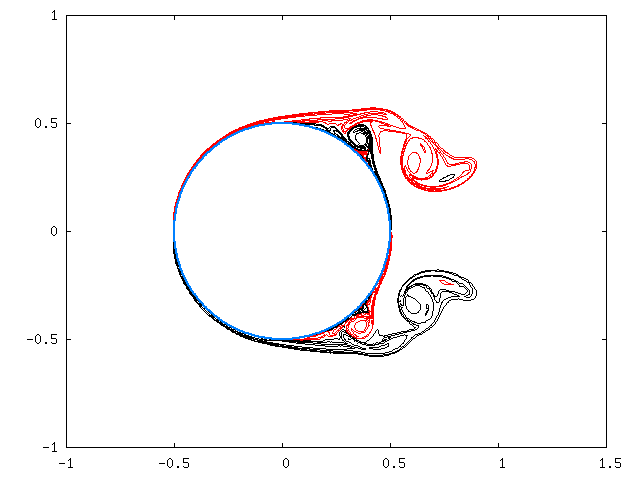}
\caption{Vorticity contours for $Re=9500$ at $t=2$: black, positive;
red, negative.  Going from the far field towards the 
cylinder, the contours are for $|\omega |$=5,10,25,50,100,200.   
The contours above the axis ($y>0$) are from the DVM code with a single 
Cartesian redistribution grid with 3000 vortex panels, $h=1/1000$, 
$\Delta t = 0.005$, 
and those below the axis from the FDS code. }
\label{cont_both_Re9500_single}
 \end{center}
\end{figure}

Figure \ref{cont_both_Re9500_single} shows vorticity contours from 
both the FDS method and the DVM calculation with the finer grid.  
There is good agreement, but not as close as with the 
body fitted grid (Figure \ref{cont_both_Re9500_t2}).
A similar comparison with the lower resolution DVM solution showed 
a similar general structure (e.g.\ the position of the large 
vortices sitting off the surface) but significant differences 
at smaller scales, reflecting a lack of resolution.  

\vspace{0.5cm}
\section{Conclusions}

A simple redistribution scheme for viscous flow has been presented.  
Unlike other redistribution schemes, it operates by redistributing 
the circulation in a vortex element to a set of fixed nodes rather than 
transferring circulation between vortex elements.  A new distribution of 
vortex elements can then be constructed from the circulation on 
the nodes. A major advantage 
of the scheme is that the solution of the redistribution problem is 
given explicitly by a set of simple algebraic equations.  A further 
advantage is that core overlap is not an issue for the viscous solution.  

The scheme will be stable provided the viscous diffusion length is less 
than the smallest mesh length in the problem.  This restriction is 
similar to that found with other redistribution schemes. 

The ability of the scheme to produce high resolution solutions has been 
demonstrated through a series of test problems.   Accurate estimates for 
both the total body and pointwise surface forces can be obtained when using a body fitted mesh 
near the surface of the body.  Accurate estimates of the total forces 
on the body can be obtained when using a non conservative scheme with the 
body embedded in a Cartesian mesh.  

The solution presented for the redistribution problem is the simplest 
possible which satisfies the equations and has the required symmetry.  It 
would be possible to obtain higher order solutions by extending the 
the computational stencil and setting higher moments to zero.  
There is no requirement to use the same mesh throughout the computational 
domain, and the use of local grid refinement is straightforward.  
Also, the method extends naturally to three-dimensions. 

\vspace{0.5cm}


\begin{thebibliography}{99}

\bibitem{Chorin}
A.J.\ Chorin.  Numerical study of slightly viscous flow.  
{\em J.\ Fluid Mech.}, {\bf 57}, 785 (1973).  

\bibitem{DMG}
P.\ Degond and S.\ Mas-Gallic.  The weighted particle method for 
convection-diffusion equations.  {\em Math.\ Comput.}, {\bf 53}, 485-526 (1989).

\bibitem{SvD}
S.\ Shankar and L.\ van Dommelen. A new diffusion procedure for vortex methods.  
{\em J.\ Comput.\ Phys.}, {\bf 127}, 88-109 (1996). 

\bibitem{Shankar}
S.\ Subramaniam.  A new mesh-free vortex method.  
PhD thesis, Florida State University, 1996. 

\bibitem{LG}
I.\ Lakkis and A.\ Ghoniem. 
A high resolution spatially adaptive vortex method for separating 
flows.  Part I: two-dimensional domains.
{\em J.\ Comp.\ Phys}, {\bf 228}, 491-515 (2009). 

\bibitem{Cottet}
G.-H.\ Cottet and P.D.\ Koumoutsakos.  
{\em Vortex Methods: Theory and Practice.}  Cambridge University Press, 
2000.  

\bibitem{KT}
K.\ Takeda, O.R.\ Tutty and A.D.\ Fitt.    
A comparison of four viscous models for the discrete vortex method. 
AIAA 13th Computational 
Fluid Dynamics Meeting, Colorado, July 1997.  AIAA paper 97-1977, 
11pp.


\bibitem{SS}
P.A.\ Smith and P.K.\ Stansby.  
Impulsively started flow around a circular cylinder by the 
vortex method.
{\em J.\ Fluid Mech.}, {\bf 194}, 45-77 (1988).

\bibitem{CT}
N.R.\ Clarke and O.R.\ Tutty.  
Construction and Validation of a discrete vortex method for the 
two-dimensional incompressible Navier-Stokes equations.  
{\em Computers \& Fluids}, {\bf 23}, 751-783 (1994).  

\bibitem{KL}
P.\ Koumoutsakos and A.\ Leonard.  
High-resolution simulations of the flow around an 
impulsively started cylinder using vortex methods. 
{\em J.\ Fluid Mech.}, {\bf 296}, 1-38 (1995).

\bibitem{simplex}
B.D.\ Bunday. 
{\em Basic Linear Programming}, Edward Arnold (London) (1984).


\bibitem{Spalart}
P.R.\ Spalart.  
Vortex methods for separated flows.
Von Karman Inst.\ for Fluid Mechanics, 
Lecture Series 1988-05 (1988).

\bibitem{Wu}
J.C.\ Wu. Theory for aerodynamic force and moment in viscous flow.
{\em AIAA J.}, {\bf 19}, 432-441 (1981).

\bibitem{PW} 
P.\ Ploumhans and G.S.\ Winckelmans. 
Vortex methods for high-resolution simulations of viscous flow 
past bluff bodies of general geometry.
{\em J.\ Comp. Phys}, {\bf 165}, 354-406 (2000).

 
\bibitem{BLY}
M.\ Bar-Lev and H.T.\ Yang.
Initial flow field over an impulsively started circular cylinder.  
{\em J.\ Fluid Mech.}, {\bf 72}, 625-647  (1975).

\bibitem{SCRD}
S.C.R.\ Dennis and S.\ Kocabiyik.
An asymptotic matching condition for unsteady boundary-layer 
flows governed by the Navier-Stokes equations.
{\em IMA J.\ Appl. Maths}, {\bf 47}, 81-98 (1991).

\bibitem{Moffat}
H.K.\ Moffat.
Viscous and restive eddies near a sharp corner.
{\em J.\ Fluid Mech.}, {\bf 18}, 1-18 (1964).

\end{thebibliography}
\end{document}